\documentstyle{mn}
\newif\ifAMStwofonts

\def\lsim{\lower.5ex\hbox{$\; \buildrel < \over \sim \;$}}
\def\gsim{\lower.5ex\hbox{$\; \buildrel > \over \sim \;$}}

\title{Helium preenrichment in the first stars}
\author[L. Chuzhoy]
       {Leonid Chuzhoy \\
        McDonald Observatory and Department of Astronomy, The University of Texas at Austin, RLM 16.206, Austin, TX 78712, USA\\
	email: chuzhoy@astro.as.utexas.edu}

\pagerange{\pageref{firstpage}--\pageref{lastpage}}
\begin{document}
\maketitle

\begin{abstract}
We show that element diffusion can produce large fluctuations in the initial helium abundance of stars. Diffusion time-scale, which in stellar cores is much larger than the Hubble time, can fall below $10^8$ years in the neutral gas clouds of stellar mass, dominated by collisionless dark matter or with dynamically important radiation or magnetic pressure. Helium diffusion may therefore explain the recent observations of globular clusters, which are inconsistent with initially homogeneous helium distribution.
\end{abstract}


\section{\label{Int}Introduction}

The standard cosmological theory predicts that the epoch of nucleosynthesis ended with extremely homogeneous element distribution.  Presently observed local abundance variations are commonly attributed to stellar nuclear reactions and supernovae explosions. Yet such variations can be produced by another process, namely element diffusion.
The diffusion of elements in stars has been widely studied (e.g. Noerdlinger 1977; Proffitt \& Michaud 1991; Chaboyer et al. 1992; Richard et al. 2002).
In stellar cores, where the diffusion time-scale is typically of order $10^{11}$ years, diffusion can increase helium abundance by up to a few per cent, thereby accelerating their evolution.
However, as we show in this paper, diffusion can make much stronger impact on helium abundance of the diffuse gas clouds of stellar mass. Therefore, stars produced by collapse of such clouds, will have high helium abundance from the moment of their formation.

Element diffusion is caused by pressure gradients which impart different accelerations to elements with different atomic mass. For diffusion to be fast, low friction (collision rate) between the particles has to be accompanied by large pressure gradient.  However, to sustain pressure gradient for sufficiently long time requires an opposing force, such as gravity. In a system composed by gas only, increasing gravitational force is possible only by increasing density, which would simultaneously increase friction. One way out to avoid this problem is provided by the collisionless dark matter (DM), which contributes to gravitational force without increasing friction. 
 Efficient diffusion is also possible in systems with dynamically important magnetic or radiation pressure. There strong gradients in the individual pressure components (magnetic, radiation, thermal) may exist even when the total pressure is constant, thus allowing to avoid strong gravity/high density requirement. Finally, a more exotic alternative for efficient diffusion comes from the MOND model (Milgrom 1983; Bekenstein 2004), which predicts stronger gravitational force for the low-mass rarefied clouds than the standard newtonian theory.

We have found that prior to the end of reionization, diffusion should have produced many helium rich gas clouds of stellar mass that could make up to from several per cent (in the standard $\Lambda$CDM model) and to half of the total baryon budget (in the MOND model). Later, as the gas is accreted by small halos and recombines, diffusion may again become efficient on the stellar mass scales, until the ambient pressure and metallicity rise to sufficiently high values.
Helium diffusion, therefore, may explain large fluctuations of primordial helium abundance inferred from the recent observations of globular clusters (Bedin et al. 2004; Norris 2004; Lee et al. 2005; Piotto et al. 2005).

In \S 2, 3 and 4 we study the diffusion of helium, respectively, in the DM dominated minihalos, in the MOND model and inside pressure-confined clouds. 
In \S 5 we consider the effect of turbulence on the helium abundance variation.
In \S 6 we discuss the implications of our results.

\section{Gravitational sedimentation inside dark matter minihalos}
In absence of radiation and magnetic pressure, the dynamics of different atomic species is determined by gravitational field, partial pressure gradients and frictional forces due to collisions:
\begin{eqnarray}
\label{acc}
a_{\rm i}=g-\frac{\nabla (n_{\rm i} kT)}{n_{\rm i}m_{\rm i}}+\Sigma \frac{V_{\rm j}-V_{\rm i}}{\tau_{\rm ij}},
\end{eqnarray}
where  $V_{\rm i}$, $a_{\rm i}$, $n_{\rm i}$ and $m_{\rm i}$ are respectively the average velocity, acceleration, number density and particle mass of the species i, and $\tau_{\rm ij}$ is the friction coefficient between species i and j. 
Since $\tau_{\rm ij}$, which is in effect the time-scale for the particle belonging to species i to be stopped by collisions with species j, is typically much smaller than all other relevant time-scales, the diffusion can be described as a steady state process with $a_{\rm i}=0$ in the gas central mass frame. Furthermore, when the abundances are still nearly homogeneous, we can take  $\nabla (n_{\rm i} kT)/n_{\rm i}=\mu\nabla p/\rho$, where $p$ and $\rho$ are, respectively,  the total gas pressure and density, and $\mu$ is the mean molecular weight. Using this in eq. (\ref{acc}) we get
\begin{eqnarray}
\label{vel1}
\sum_{\rm j} \frac{V_{\rm j}-V_{\rm i}}{\tau_{\rm ij}}=
\mu m_{\rm H}/m_{\rm i}\frac{\nabla p}{\rho}-g.
\end{eqnarray}
For maxwellian velocity distribution $\tau_{\rm ij}$ is given by
\begin{eqnarray}
\label{tauc}
 \tau_{\rm ij}=\frac{3\sqrt{2\pi}}{16n_{\rm j}\sigma_{\rm ij}}\sqrt{\frac{m_{\rm i}(m_{\rm i}+m_{\rm j})}{m_{\rm j}kT}},
\end{eqnarray}
where $\sigma_{\rm ij}$ is the momentum transfer cross-section.
For the scattering of neutral hydrogen and helium atoms we adopt the experimental value, $\sigma_{\rm HeH}=0.9\times 10^{-15}{\rm cm^2}$ (Khouw, Morgan \& Schiff 1968). 
For the ionized gas the cross-sections are typically much larger, so in this paper we restrict the calculations to the neutral medium.

For hydrostatic equilibrium, $\nabla p/\rho=g$, and the standard abundances ($X=0.75, Y=0.25$), the relative velocity of helium and hydrogen atoms is
\begin{eqnarray}
\label{VHe}
 V_{\rm He}-V_{\rm H}= 80{\rm\; m/s}\times \hspace{2cm} \nonumber \\
 \left(\frac{g}{10^{-12}{\rm m/s^2}}\right) \left(\frac{n}{10^3{\rm m^{-3}}}\right)^{-1} \left(\frac{T}{10{\rm K}}\right)^{-1/2}.
\end{eqnarray}
Assuming for simplicity the spherical symmetry we can derive the evolution of particles abundances from the continuity equation
\begin{eqnarray}
\label{cont}
 \partial_t n_{\rm i}=-r^{-2}\nabla(r^2 n_{\rm i} V_{\rm i}),
\end{eqnarray}

For systems with uniform temperatures combining eqs. (\ref{VHe}) and (\ref{cont}) and
replacing $g$ with $-GM/r^2$, where $M$ is the total mass inside radius $r$, gives the characteristic time-scale for helium enrichment
\begin{eqnarray}
\label{timeHe}
\tau_{\rm He}=\partial_t \ln(n_{\rm He}/n_{\rm H})^{-1}= 10^{11} f_{\rm g}T^{1/2}{\rm yrs},
\end{eqnarray}
where $f_{\rm g}=\rho_{\rm gas}/(\rho_{\rm gas}+\rho_{\rm DM})$ is the local gas mass fraction. Conveniently $\tau_{\rm He}$ is independent of both mass and the radius of the system, but is a function of temperature and the gas fraction only.

When $\tau_{\rm He}$ is much larger than the age of the halo the helium abundance in the minihalos grows linearly with time
\begin{eqnarray}
\label{dY}
\frac{\Delta Y}{Y}=\frac{t}{\tau_{\rm He}}=6\cdot 10^{-3} f_{\rm g}^{-1}T^{-1/2}\left(\frac{1+z}{10}\right)^{-3/2}\times \nonumber \\
\left[1-\left(\frac{1+z}{1+z_{\rm f}}\right)^{3/2}\right]\left(\frac{\Omega_{\rm m,0}h^2}{0.15}\right)^{-1/2}, \hspace{1cm}
\end{eqnarray}
where  $z_{\rm f}$ is the formation redshift of the halo and 
\begin{eqnarray}
t=t_{\rm Hubble}\left[1-\left(\frac{1+z}{1+z_{\rm f}}\right)^{3/2}\right]= 5.4\cdot 10^8\;{\rm yrs}\times \hspace{1cm}\nonumber \\
\left(\frac{1+z}{10}\right)^{-3/2}\left[1-\left(\frac{1+z}{1+z_{\rm f}}\right)^{3/2}\right]\left(\frac{\Omega_{\rm m,0}h^2}{0.15}\right)^{-1/2}
\end{eqnarray}
is its age at redshift $z$.
Inside large halos whose virial temperature, $T_{\rm vir}$, is far above the background gas temperature, $T_{\rm bg}$, the gas fraction is close to the mean cosmic value $\sim 0.15$, so that $\Delta Y/Y\ll 1$. On the other hand in minihalos with $T_{\rm vir}<T_{\rm bg}$ the gas accretion is suppressed by pressure. At virialization the dark matter density contrast, whose evolution is nearly independent of gas dynamics, reaches $\sim 200$, so that the gas fraction drops roughly to $0.15/200\sim 10^{-3}$. Moreover the gas expansion continues until $T_{\rm bg}$ falls below $T_{\rm vir}$, thus further lowering $f_{\rm g}$.

After the gas thermally decouples from radiation at $z\sim 150$, its temperature drops adiabatically (Peebles 1993).
\begin{eqnarray}
\label{Tbg}
T_{\rm bg}=0.017(1+z)^2.
\end{eqnarray}
Later between $z\sim 10$ and $z\sim 20$ the gas is reheated by X-rays and the UV photons, followed by reionization. From eq. (\ref{timeHe}) we see that the optimal conditions for diffusion in DM minihalos are during the epoch when the temperature drops to its minimum, which is roughly between 2 and 10 K. At these temperatures and for $f_{\rm g}=10^{-3}$, the change in helium abundance, $\Delta Y/Y$, is of order unity (eq. (\ref{dY})).

It should be noted, however, that $\Delta Y/Y$ can not be large when $T_{\rm vir}\ll T_{\rm bg}$. The approximation we used so far for all elements, $\mu m_{\rm p} g=\nabla (n_{\rm i}kT)/n_{\rm i}$, will not be valid after diffusion had enough time to make $\Delta Y/Y$ close to its equilibrium value (around $(m_{\rm He}/m_{\rm H}-1)T_{\rm vir}/\mu T_{\rm bg}=2.5T_{\rm vir}/T_{\rm bg}$ for $T_{\rm vir}\ll T_{\rm bg}$). So $\Delta Y/Y$  can be of order unity only when $T_{\rm vir}\sim T_{\rm bg}$.

The condition $T_{\rm vir}\sim T_{\rm bg}$, required for efficient diffusion, sets a characteristic scale for the individual gas cloud that can become helium rich. The virial temperature of a halo of mass $M$ and density $\rho$ is 
\begin{eqnarray}
\label{temv}
T_{\rm vir}= 
15 \;{\rm K}\left(\frac{M}{10^3 M_{\odot}}\right)^{2/3}\left(\frac{1+z_{\rm f}}{10}\right). \hspace{2cm}
\end{eqnarray}
Setting $T_{\rm vir}\sim T_{\rm bg}$ and using eq. (\ref{dY}), we find the mass of the gas cloud whose helium fraction is increased by $\Delta Y/Y$ 
\begin{eqnarray}
\label{mlim}
M_{\rm cl}=f_{\rm g}M=0.2 M_\odot \left(\frac{\Delta Y}{Y}\right)^{-1} \left(\frac{1+z}{10}\right)^{-1} \left(\frac{1+z}{1+z_{\rm f}}\right)^{3/2}\times \nonumber \\
\left[1-\left(\frac{1+z}{1+z_{\rm f}}\right)^{3/2}\right]\left(\frac{\Omega_{\rm m,0}h^2}{0.15}\right)^{-1/2}. \hspace{2cm}
\end{eqnarray}
Thus the increase of order unity in helium abundance can be achieved in the gas clouds of stellar mass. However, since the efficient diffusion is restricted to the regions with low gas fraction, prior to reionization no more than a few per cent of the baryons can reside in the helium rich clouds. In regions, whose evolution is still in the linear regime, the variation of helium abundance does not exceed 0.1-1 \% (Medvigy \& Loeb 2001).

To verify that the assumption of hydrostatic equilibrium is a valid approximation, we need to check that the sound crossing time, $t_{\rm s}$, in a cloud is smaller than its dynamical time, $t_{\rm dyn}=(G\rho)^{-1/2}$, while both of them are smaller than the Hubble time. The second condition, $t_{\rm dyn}< t_{\rm Hubble}$, is always satisfied inside the virialized dark matter halos, since  $t_{\rm dyn}/t_{\rm Hubble}=(8\pi/3\delta_{\rm dm})^{1/2}\sim 0.2$, where $\delta_{\rm dm}\sim 200$ is the dark matter density contrast. The ratio $t_{\rm s}/t_{\rm dyn}$ in the adiabatically compressed gas clouds is
\begin{eqnarray}
\frac{t_{\rm s}}{t_{\rm dyn}}=0.6  (1+\delta_{\rm gas})^{-2/3} \left(\frac{0.2M}{M_{\odot}}\right)^{1/3}\left(\frac{1+z}{10}\right)^{-1/2}\times \nonumber \\
\left(\frac{1+\delta_{\rm dm}}{200}\right)^{1/2}\left(\frac{\Omega_{\rm b,0}h^2}{0.02}\right)^{-1/3}, \hspace{3cm}
\end{eqnarray}
where $\delta_{\rm gas}$ is the gas density contrast. Thus inside the stellar mass clouds the hydrostatic equilibrium is indeed a reasonable approximation.

\section{Diffusion in MOND model}
Modified gravity has been suggested as an alternative to the collisionless dark matter (Milgrom 1983; Bekenstein 2004). In MOND theory in the limit of weak gravity, gravitational acceleration is given by $g=\sqrt{g_{\rm N}g_0}$, where $g_{\rm N}=GM/r^2$ is the standard Newtonian value. Based on the galactic rotation curves the value of $g_0$ is expected to be around $10^{-10}\; {\rm m\cdot s^{-2}}$.
 Using this in eqs. (\ref{VHe}) and (\ref{cont}) we derive the new diffusion time-scale 
\begin{eqnarray}
\label{timeHe2}
\tau_{\rm He}= 10^{11} T^{1/2} (\frac{g_{\rm N}}{g_0})^{1/2} {\rm yrs}= 0.8 \cdot 10^8  {\rm yrs}\times  \hspace{2cm} \nonumber \\
 T^{1/2}(1+\delta_{\rm gas})^{1/3}\left(\frac{0.2M}{M_\odot}\right)^{1/6}\left(\frac{1+z}{10}\right)\left(\frac{\Omega_b h^2}{0.02}\right)^{1/3}.
\end{eqnarray}
The dynamical time for modified gravity is $t_{\rm dyn}=(G\rho g/g_{\rm N})^{-1/2}$, so the ratio $t_{\rm s}/t_{\rm dyn}$ in this case is
\begin{eqnarray}
\frac{t_{\rm s}}{t_{\rm dyn}}=0.4  (1+\delta_{\rm gas})^{-1/3} \left(\frac{0.2M}{M_{\odot}}\right)^{1/4}\left(\frac{1+z}{10}\right)^{-1},
\end{eqnarray}
showing that the assumption of hydrostatic equilibrium is again reasonable for stellar mass clouds.

Unlike the case with newtonian gravity, in MOND $\tau_{\rm He}$ is no longer scale independent.
The cloud virial temperature in the MOND regime ($g\ll g_0$) is $T_{\rm vir}\approx  10\;{\rm K} (M/M_{\odot})^{1/2}$, so prior to reionization, when $T_{\rm bg}$ falls to a few kelvins, the diffusion is again most efficient in clouds of stellar mass. However, in MOND all gas may undergo efficient diffusion (i.e., $\tau_{\rm He}\lsim t_{\rm Hubble}$) and not just a small fraction that resides within dark matter minihalos with $M\sim M_{\rm Jeans}$.

\section{Diffusion in pressure-confined clouds}
Radiation pressure from the resonant Ly$\alpha$ photons is capable of supporting large gas clouds  (Braun \& Dekel 1989). Since hydrogen resonance photons are far more abundant than helium resonance photons, the radiation pressure exerts stronger force on hydrogen than helium atoms. The magnetic pressure presents a similar case.  The magnetic field acts directly only on charged particles (i.e., electrons and ions). Those in turn collide with the neutral atoms, transferring to them the acquired momentum. Since the collisional cross-section of ions with hydrogen atoms is several times larger than with helium atoms (Osterbrock 1961; Krstic \& Schultz 1999), the magnetic pressure also exerts stronger force on hydrogen atoms. The eq. (\ref{vel1}) can thus be expanded to account for all pressure components
\begin{eqnarray}
\label{V2}
\frac{V_{\rm He}-V_{\rm H}}{\tau_{\rm HeH}}=\frac{\nabla(\alpha_{\rm th}p_{\rm th}+\alpha_{\rm rad}p_{\rm rad}+\alpha_{\rm B}p_{\rm B})}{\rho},
\end{eqnarray}
where $p_{\rm th}$, $p_{\rm rad}$ and $p_{\rm B}$ are, respectively, thermal, radiation and magnetic pressure, $\alpha_{\rm th}=1-\mu m_{\rm p}/m_{\rm He}\approx 0.7$, and $\alpha_{\rm rad}$ and $\alpha_{\rm B}$ are typically close to unity.
When the total pressure, $p=p_{\rm th}+p_{\rm rad}+p_{\rm B}$, is constant, but radiation or magnetic pressure have strong variation on the scale $R$, eq. (\ref{V2}) becomes
\begin{eqnarray}
\label{V3}
\frac{V_{\rm He}-V_{\rm H}}{\tau_{\rm HeH}}\approx \frac{(1-\alpha_{\rm th}) p}{\rho R}\approx \frac{0.3 p}{\rho R},
\end{eqnarray}
The time-scale for helium enrichment is then
\begin{eqnarray}
\label{timeHe3}
\tau_{\rm He}=\frac{R}{V_{\rm He}-V_{\rm H}}=1.7\cdot 10^8 {\rm yrs} \times \hspace{2cm} \nonumber\\ 
 \left(\frac{M_{\rm cl}}{0.2M_\odot}\right)^{2/3} \left(\frac{p/k}{100\;{\rm K\cdot cm^{-3}}}\right)^{1/3} \left(\frac{T}{10^4\;{\rm K}}\right)^{-5/6}.
\end{eqnarray}
From the above equation it follows that when pressure is fixed, the diffusion is most efficient at the highest temperature possible for the neutral gas ($T\sim 10^4$ K). 
In the interstellar medium of the Milky Way, where the metallicity and pressure are high ($p/k\sim 2\cdot 10^4\;{\rm K\cdot cm^{-3}}$), it generally takes around a few million years for a cloud with $T\sim 10^4$ K either to cool down and collapse into a star, or to be destroyed through shockheating or ionization. Since for a given pressure, $\tau_{\rm He}$ is of order  $10^9$ years, the expected helium variation due to diffusion in stars that currently form in the Milky Way, typically should not exceed one per cent. By contrast, in the first halos with $T_{\rm vir}\sim 10^4$ K, which form out of the metal-poor gas, and where the pressure is significantly lower ($p/k\sim 10^2-10^3\;{\rm K\cdot cm^{-3}}$), the typical life-time of a cloud may be comparable to $\tau_{\rm He}$. Therefore, significant variation of initial helium abundances may be possible for the early generations of stars.

\section{Turbulent mixing}
Small-scale turbulent motion tends to reverse the process of helium segregation by mixing the gas from regions with different abundances. When helium abundance gradient, $\nabla Y$, exceeds $(V_{\rm He}-V_{\rm H})/V_{\rm turb}R_{\rm turb}$, where $V_{\rm turb}$ and $R_{\rm turb}$ are the characteristic velocity and the length scale of the turbulence, turbulent mixing prevails over helium sedimentation. In case both operate at the same time, then in a cloud of radius $R_{\rm cl}$
helium abundance may vary by at most $\Delta Y/Y=(V_{\rm He}-V_{\rm H})R_{\rm cl}/V_{\rm turb}R_{\rm turb}$. When turbulence becomes strong after helium sedimentation had some time to act unopposed, then over time the initial abundance gradient is diluted by a factor $\sim e^{-V_{\rm turb}R_{\rm turb}t/R_{\rm cl}^2}$.
By contrast, large-scale turbulent motion, with $R_{\rm cl}\ll R_{\rm turb}$, has no effect on the small-scale abundance fluctuations. Furthermore, if the cloud is magnetized, the magnetic stress may be able to suppress the turbulent mixing altogether.

In general there is no reason to expect strong turbulent motion in the low density gas clouds that were first to form out of cosmological density perturbation field (\S 2 and 3). Strong turbulence may arise later, as these clouds are accreted by larger objects and become pressure-confined, but its length scales, amplitude and duration are very uncertain. 
However, as has been pointed out by Bruston et al. (1981), while the impact of turbulence on the abundance variation in the interstellar clouds is hard to calculate, observations suggest that it should be limited. In particular turbulence failed to destroy an order of magnitude variation in the abundances of different isotopes of carbon (Encrenaz, Falgarone \& Lucas 1975; Goldsmith \& Langer 1978) and the separation between the different types of dust grains (Carrasco, Strom \& Strom 1973).
Therefore it seems likely that the fluctuations of helium abundance may survive as well.

\section{Discussion}

We have shown that diffusion can significantly increase the helium abundance of protostellar clouds. Similarly diffusion may also produce spatial variation of the deuterium (Bruston et al. 1981) and lithium abundances. The amplitude of the variation depends on many factors (magnetic field, outside pressure, turbulence etc.) and in some systems the effect of diffusion may still be negligible. However, as demonstrated by our calculations, the conservation of primordial abundances until the onset of stellar nucleosynthesis, cannot in general be taken for granted.

Recent observations strongly suggest that stars in at least some globular clusters have been formed with enhanced helium abundance (Bedin et al. 2004; Norris 2004; Lee et al. 2005; Piotto et al. 2005). It has been even suggested that the initial abundance of helium may be the missing ``second parameter'' (Caloi \& D'Antona 2005). Several scenarios involving early pollution by helium-rich but metal-poor stars have been suggested as an explanation. However, Bekki \& Norris (2005) have shown that such pollution is very unlikely, unless the helium rich gas comes from the outside source and is kept in place by high external pressure. Therefore, diffusion provides a strong alternative as an explanation to high helium abundance.

\section*{Acknowledgment}
The author thanks Volker Bromm, Abraham Loeb, Adi Nusser and Marcelo Alvarez for stimulating discussions, and  McDonald Observatory for the W.J. McDonald Fellowship.

\end{document}